# Virtual Learning Environments – A Survey

[1]SNEHA J.M, [2]G.S.NAGARAJA

[1] M.Tech Student, Department of Computer Science & Engg, R V College of Engineering
[2] Associate Professor, Department of Computer Science & Engg, R V College of Engineering

**Abstract ---** This paper is based on the study of existing literature, highlights the current state of the work proposed to implement technically enhanced learning.

Technology developments and network infrastructure improvements, (specifically the world wide web) are providing exciting opportunities for the use of computers in all areas. These developments have fit together with an evolving role for education as more students wish to study at a distance, part-time, or wish to integrate their education with their professional career. With the market becoming increasingly 'mature', e-learning has almost become a major plank in both national and institutional strategies.

At the same time, virtual learning system (vls) is also gaining its popularity among its users. It has brought in a great revolution in itself. In the advanced learning strategy, virtual learning systems depends on level and sector of working, usage of functions, purpose of usage, required online resources to perform computationally intensive operations such as information sharing and collaborative work. Often the institutions prefer the latest and the best technology which is cost effective and provide the best features which meet up all requirements. Virtual learning environment enables operator or professional to bring together in one place a variety of prevailing resources, suchlike tasks and formative feedback, and links to law reports, statutes and journal articles, all intended at summing up value to student learning – managing learning experience without the burden of communication and providing successful delivery of education and training with flexibility.

In spite of the hype achieved by organizations in techanically enhanced learning, the growth of the virtual learning systems users is still below expectations due to the risks associated with the implementation strategy and provision of technical support.

## I. INTRODUCTION

In order to understand working of Virtual Learning Systems, it is necessary to define the concept of Virtual Learning.

The term 'virtual learning environment' (VLE) refers to the components in which learners and tutors participate in online interactions of several kinds, comprising online learning.

E-learning refers to the use of electronic media and information and communication technologies (ICT) in education. E-learning is widely enclosure of all kinds of educational technology in learning and teaching. E-learning is enclosure of, and is widely synonymous with technology-enhanced learning (TEL), computer-based training (CBT), multimedia learning,computer-assisted instruction, computer-based instruction (CBI) or internet-based training (IBT), web-based training (WBT), online education, computer-aided instruction (CAI), virtual education, m-learning, virtual learning environments (VLE) (which are also called learning platforms), and digital educational collaboration. These other alternative names dwell on a specific component, aspect or delivery method.

A Virtual Learning Environment (VLE) is a system for transfering learning materials to students by means of the web. These systems constitute student tracking, communication tools, assessment and collaboration. They can be accessed both on and off-campus, which means that they can encourage students' learning outside lecture hall 24 hours a day, seven days a week. This facilitates institutions to teach not only traditional full-time students but also those who cannot regularly visit the campus due to geographic or time restrictions, i.e. those doing evening classes, on distance learning courses, or workers studying part-time[1].

There are various kinds of VLE, which all operate slightly varingly but eventually accomplish the identical function and can transfer the same learning materials. A Higher Education academy is probable to have a licence for a VLE that belong to any one of the following three classifications

- Off-the-shelf, such as Blackboard or WebCT
- Open source (often free to use and adapt whereas support is charged), for instance Moodle
- Bespoke (developed by institutions for their own individual needs)

There are some international standards associated with VLEs which help to make content and assessments interoperable (i.e. they can be used in different types of VLE). The standard for content is called 'Sharable Content Object Reference Model' (SCORM) and the standard for assessments is called 'Question and Test Interoperability' (QTI)[1].

Typically, VLEs provide facilities for managing the learning experience, communicating the intended learning experience and facilitating tutors' and learners' involvement in that experience. The learning experience needs to be communicated via syllabi, complete course content or copies of visual aids/handouts, plus additional resources, links to resources in libraries and on the Internet. Easy authoring tools or standard office software used for authoring should be available to aid this. The learning experience is facilitated typically via self-assessment quizzes and communications tools such as e-mail, threaded discussions and chat rooms. To allow all of this, the systems should provide differential access rights for instructors and students (roles). All the various functions and resources need to be capable of being hyperlinked together within a consistent interface[2][3].

From the earliest Internet-based learning systems, people started exploring the possibilities of the worldwide web as a means of supporting learning. The earliest systems which satisfied the 'consensus view' of VLE elements began to appear and included systems such as WebCT and Lotus LearningSpace.

During this period, systems were developed which took a pedagogically focused view of e-learning. Several notable examples developed in the UK included Boddington (developed at Leeds University), Colloquia (developed at Bangor University) and COSE (developed by a team at Staffordshire University, of which I was leader). Systems like these were largely constructivist in their approaches, were learner-centric and used novel interfaces[4][5].

As things developed, along with the exponential uptake of the use of the web, it became clear that systems which had interfaces which looked like websites and were delivered via browsers were becoming the most successful. Interestingly, research conducted in 2004 showed that 'ease of use' – especially by teaching staff – was the prime consideration in VLE selection. Respondents made great play of the need for an 'intuitive' interface.

The 'early adopter' of VLEs largely cited reasons such as learner centeredness, pedagogic change, diversification, and coping with large numbers as reasons for getting involved. By the time of the survey conducted in 2003, leading reasons for selection were ease of use, cost, flexibility and, interestingly, level of use by other institutions[5].

By the time of the first JISC/UCISA 'MLE Landscape' survey in 2005, the vast majority of institutions that responded were using a VLE and cited the drivers as:

- Enhancing access to learning for students off campus
- Broadening participation/inclusiveness
- Upgrading the nature of teaching and learning
- Utilizing technology to deliver e-learning
- Student expectations
- Reforming access for part-time student.

## II. . SURVEY OF EXISTING VIRTUAL LEARNING SYSTEMS

We present the notable systemss that have been proposed in the scientific journals and conferences concerning to technical interoperability, flexibility and cost-effectiveness with respect to deployment. We describe the features, technology and scenarios of use of a number of VLEs that are currently available.

### A. Webct (Course Tools) or Blackboard Learning System

According to the team that developed WebCT it is " a tool that facilitates the production of sophisticated Web-based educational environments." It can be used flexibly to establish complete online courses, or to publish materials that add existing courses. All communication with WebCT come about through a web browser.. Essentially a webCT course comprises of a series of linked HTML pages that define a path or "road-map" through the course material. The course content is supplemented by webCT tools which can be built into the course design by simply dragging the appropriate tool icon onto the web page. This creates an active link to a 'toolpage'[6].

There are three main aspects to webCT:

1. A presentation tool that allows the designer to customise the look and feel of the course pages.
2. A set of student tools that can optionally be integrated anywhere in the course.
3. A set of administrative tools to aid the tutor manage the course when it is in progress.

A webCT course centres around a single course homepage, which appears whenever a student logs on to their course. This page contains links to the various course pages and the tools. The learner tools in webCT include: asynchronous communications tools (e-mail and conferencing) and also a chat facility; student evaluation and self-evaluation tools such as online quizzes and MCQs; an image repository; a glossary database; learner collaboration and presentation areas; content annotation, homepage generation, course navigation and searching tools. Tutors can track student progress including the number of times and when they have accessed the course.

Technology: This is a client-server system with all interaction through a web browser client. Supported server platforms are currently only Unix but an NT 4.0 version is to be announced.

The proposed framework fails to meet all guidelines for accessibility. WebCT's user interface has been criticized as needlessly complex and unintuitive. The "Vista" version of the product represented an attempt to derive balance between flexibility and ease of use, however it has also been the target of ease-of-use criticisms.

Some WebCT criticisms which were apparent consist of difficulties deploying it in multiple tabs or browser windows, large dependence on Java for its user experience,[6] practice of too many browser framesets, concerns with some features requiring pop-up blockers to be turned off, and difficulties using standard browser navigation tools (i.e. the Back and Forward commands).

B. Virtual-U

Virtual-University[7] like the other systems described is a server based integrated environment for education that uses a web client. One of the major differences between Virtual-U and other systems is the use of a campus metaphor within which locations and objects are used to represent the different tools and activities. The homepage contains a map of the Virtual_U campus which consists of :

- A Course Room – containing details of the courses
- A Library – containing resources and links to library pages and search engines
- A Gallery – for presentation of multimedia resources
- A Conference Building – from which conferences can be accessed
- A Café – another conference space for more casual student or staff interaction
- A Personal Workspace – The workspace provides links to all other areas of the campus and also includes a number of tools for managing your own work. These tools include a calendar for personal time planning, a submissions box for viewing assignments, a grade-book allowing students to view their own performance or for instructors to monitor student progress, a glossary and preference options.

Additionally, there is also a detailed help system providing course designers with pedagogical guidelines for the use of the system.

Technology: Web browser client, Server software runs on Unix or NT.

Cons:

- Interworking with secure IT environments
- Greater cost of support, customization, and installation
- Larger reliance on user networks, communities
- Lock users out of capabilities not designated in their user profile
- Protect against unauthorized login

C. .LRN

.LRN[8] is a fully internationalized open source portal and application framework framed to assist online collaborative learning communities and blended learning environments. .LRN is grounded on the principle that learning is a social experience (Wegerif), that effective learning usually carries out in the context of communities (Alavi; Wenger ), and that administration of these communities should be distributed. Each learning community has its own stakeholders and needs to define its own unique set of communications, so the software is sketched to be adaptable and permit delegation of administrative roles as close to the learners as possible. While some e-learning environments are created around a course catalog (course management systems, CMS) and other are constructed around a content management system (learning content management systems, LCMS) .LRN focuses on online communities (learning community system, LCS), with course management and content management applications as an added value.

D. Moodle

Martin Dougiamas, an Australian graduate student, originally developed Moodle in 1999 as a course management system. The platform was released to the public in 2002, initially with only the education market in mind[9].

Moodle's source code is written in PHP, a common, free scripting language that wasoriginally developed for building dynamic Web pages. The Moodle Trust manages the platform's core development, but the software is depicted to be highly modular, and numerous developers and organizations have fabricated plug-ins and other add-ons to enlarge functionality over the years. Much of Moodle's popularity rests on its ease of use. As an LMS, it provides a robust toolset, particularly thanks to plug-in modules that greatly increase its functionality. Moodle's feature set includes:

- A variety of user management options, including multiple authentication options, online profile building, and role-based assignments and permissions;
- Site administration and administrator tools;
- Course management and communications options, including chat, forums, wikis, assessment builders;
- Registration and enrollment tools and plug-ins; and more.

Despite all these features, Moodle's core design is meant to be as simple and efficient as possible. This strategy has paid dividends in the form of high satisfaction ratings among Moodle users.

Finally, Moodle lacks a full-featured competency development and management toolset, which is required by many large corporate clients. Although there are workarounds that can allow some competency tracking and reporting, Moodle is hurt in certain markets where robust competency management is a requirement.

E. Sakai

Sakai[10] is a community of commercial organizations, academic academy and individuals who work together to frame a common Collaboration and Learning Environment (CLE). The Sakai CLE is a free, community source, educational software platform allocated under the Educational Community License (a type of open source license). The Sakai Project's software is a Java-based, service-oriented application suite that is brought about to be interoperable, scalable, extensible and reliable.

In contrast to Moodle, Sakai was established on a more centrally planned model. Endowed by a Mellon Foundation grant, Sakai was framed by a consortium of five large U.S. universities, including Michigan, Stanford, MIT, UC Berkeley, and Indiana. It was based on existing tools contributed by each of the founding institutions. Sakai was released to the public in 2005 and is managed today by the Sakai Foundation, which manages its project roadmap and development. The application is programmed in Java and drafted to be a service-oriented application suite.

As a newer platform, Sakai has not yet achieved the large penetration outside the higher education marketplace that Moodle has gained. Its reputation for higher-end features, scalability, and security, however, have made it popular with large universities that need a robust solution, and it is beginning to make inroads in the government and public sector markets as well.

On the negative side, Sakai's critics point out that it, like Moodle, lacks comprehensive competency profiling and management, which causes it to be not suitable for some big enterprise environments. It also can be crucial to assosiate Sakai with other enterprise software systems, such as talent management, other HR software suites, and ERP solutions. Sakai's greater complexity also makes it more challenging to install and set up than Moodle. This makes it less suitable for simple, rapid deployment projects that require an LMS.

### III. CONCLUSION

Based on the investigations made at various levels, it clearly appears to be in robust good health when the current usage and size of market is considered, but there is a worrying trend towards bloated and monolithic systems with endless features being bolted onto them and features for disabled people need to be added.

In the future, corporately initiated or owned processes will be best dealt with by interoperating 'suits our needs best' systems, and increasingly, student initiated processes will be done on the web using their choice of tool and services. Connecting these will be the major task.

The prime role of the VLE is likely to be in articulating the intended learning experience and, to some extent, managing it. Such a system will be much 'slimmer' than most current VLEs, and whether it will still be considered to be a VLE remains to be seen. Institutions need to be reconsidering their strategies and policies now, to be ready for the changes ahead.